\documentclass[10pt,a4papper]{article}
\usepackage[utf8]{inputenc}
\usepackage{indentfirst}
\usepackage[]{graphicx}
\usepackage{geometry} 
\begin{document}
	\title{\textbf{The decays $\tau \to [\omega(782), \phi(1020)] K^{-} \nu_{\tau}$ in the extended NJL model}}
	\author{M. K. Volkov\footnote{volkov@theor.jinr.ru}, A. A. Pivovarov\footnote{tex$\_$k@mail.ru}, K. Nurlan\footnote{nurlan.qanat@mail.ru}\\
		\small
		\emph{Bogolubov Laboratory of Theoretical Physics, Joint Institute for Nuclear Research, Dubna, 141980, Russia}}
	\date{}
	\maketitle
	\small
	
\begin{abstract}
	In the extended Nambu--Jona-Lasinio model, the decay widths of the processes $\tau \to [\omega(782), \phi(1020)] K^{-} \nu_{\tau}$ were calculated. The intermediate channels with axial vector ($K_{1}(1270)$, $K_{1}(1400)$ and $K_{1}(1650)$), vector ($K^{*}(892)$ and $K^{*}(1410)$) and pseudoscalar ($K$ and $K(1460)$) mesons were considered. The results for these processes are in satisfactory agreement with the experimental data.
\end{abstract}

\large
\section{Introduction}
	The processes $\tau \to [\omega(782), \phi(1020)] K^{-} \nu_{\tau}$ were intensively investigated from both the experimental \cite{Arms:2005qg,Inami:2006vd,Aubert:2007mh} and theoretical \cite{Guo:2008sh,Dai:2018thd} points of view. In the theoretical works, the resonance chiral theory and angular momentum algebra were applied.
	
	In the present paper, the pointed processes are considered in the framework of the extended Nambu--Jona-Lasinio model (NJL) \cite{Volkov:1996br,Volkov:1996fk,Volkov:1997dd,Volkov:2005kw,Volkov:2017arr}. This model allows one to describe the scalar, psudoscalar, vector and axial vector meson nonets in both the ground and the first radially excited states in the framework of the U(3) $\times$ U(3) chiral symmetry. It has turned out to be very useful for calculation of numerous modes of the $\tau$-lepton decays because in the intermediate states of these decays the ground and the first radially excited mesons give the main contribution \cite{Volkov:2017arr}. Indeed, in the framework of this model the decays $\tau \to \pi \pi \nu_{\tau}$, $\tau \to \pi [\eta, \eta'(958)] \nu_{\tau}$, $\tau \to \pi K \nu_{\tau}$, $\tau \to [\eta, \eta'(958)] K \nu_{\tau}$, $\tau \to K K \nu_{\tau}$ \cite{Volkov:2017arr} were described without any additional arbitrary parameters. These processes have two channels. One of them is the contact channel where the $W$-boson produces the final mesons directly. The other channel includes the intermediate vector mesons in the ground and the first radially excited states. The decay $\tau \to f_{1}(1285) \pi \nu_{\tau}$ with only axial vector meson in the intermediate state was calculated \cite{Volkov:2018fyv}. The series of decays with vector and pseudoscalar mesons in the final states were also described, for example, the decay $\tau \to \pi \omega(782) \nu_{\tau}$ \cite{Volkov:2017arr} including only the contact and vector intermediate channel and the decays $\tau \to \pi [\rho(770), \rho(1450)] \nu_{\tau}$ \cite{Volkov:2019ozz} including the contact, axial vector and pseudoscalar channel. However, similar processes with strange particles are of particular interest because all the four channels are included in them.
	
	Typical examples of these decays are the processes $\tau \to [\omega(782), \phi(1020)] K^{-} \nu_{\tau}$. In the present paper, their full and differential decay widths are calculated. The processes include the contact channel and three intermediate channels: axial vector, vector and pseudoscalar ones. The nature of the intermediate meson states is discussable. For example, the axial vector state $K_{1}(1270)$ can be described in the chiral unitary approach with two poles at 1195 and 1284 MeV \cite{Roca:2005nm,Wang:2019mph}. However, in present work, we use the extended NJL model, which describes only $q\bar{q}$ resonances. The NJL model includes two strange axial vector mesons with quantum numbers $J^{PC} = 1^{++}$: the meson $K_{1A}$ and its first radially excited state considered as the physical state $K_{1}(1650)$. However, there is the strange axial vector meson $K_{1B}$ with quantum numbers $J^{PC} = 1^{+-}$ which is not described by the NJL model. The transitions between the mesons $K_{1A}$ and $K_{1B}$ are not equal to zero due to the difference between the masses of $u$ and $s$ quarks. The mixing of these mesons leads to the physical states $K_{1}(1270)$ and $K_{1}(1400)$\cite{Volkov:1986zb,Volkov:1984fr,Suzuki:1993yc}:
	\begin{eqnarray}
	\label{K1AK1B}
	K_{1A} & = & K_{1}(1270)\sin{\alpha} + K_{1}(1400)\cos{\alpha}, \nonumber\\
	K_{1B} & = & K_{1}(1270)\cos{\alpha} - K_{1}(1400)\sin{\alpha}.
	\end{eqnarray}
	
	The similar representation was applied in works of many other authors (for example, \cite{Guo:2008sh,Cheng:2017pcq,Kang:2018jzg} and other). 
	
	The contributions from the meson $K_{1B}$ to tau decays are proportional to the $U(3)$ symmetry breaking effects \cite{Suzuki:1993yc}. These effects also restrict the precision of the NJL model. Specifically, the precision of this model is determined on the basis of partial conservation of the axial current (PCAC). In the case of $U(3)$ symmetry, it can be determined with the ratio $\frac{M_{K}^{2}}{M_{\Sigma}^{2}} \approx 17\%$ \cite{Vainshtein:1970zm}. There are a large number of other sources of uncertainties. Therefore, we use our previous results to estimate the error of the model. Without any exotic states the everage uncertainty can be estimated at the level of 10\%. This error obtained from the real calculation covers all possible sources including PCAC and can be absorbed by it. Therefore, we estimate the uncertainty of this model at the level of 17\%. Thus, the direct contribution from the meson $K_{1B}$ to tau decays is within the model error and can be neglected. However, we consider the indirect influence of this meson, which leads to splitting of $K_{1A}$ into $K_{1}(1270)$ and $K_{1}(1400)$. Therefore, in the considered processes in the axial vector channels, we can only take into account the mesons described by the extended NJL model: the meson $K_{1A}$ splitted into two physical states and the radially axcited meson $K_{1}(1650)$.
	
\section{The interaction Lagrangian of the extended NJL model}	
	The fragment of the quark-meson interaction Lagrangian for the mesons included in our processes obtained from the original four-quark interactions takes the form \cite{Volkov:1996fk,Volkov:2017arr}:
	\begin{eqnarray}
	\label{Lagrangian}
	\Delta L_{int} & = &
	\bar{q} \left[ \frac{1}{2} \gamma^{\mu}\gamma^{5} \sum_{j=\pm} \lambda_{j}^{K} \left(A_{K_{1}}K^{j}_{1\mu} + B_{K_{1}}K^{'j}_{1\mu}\right) \right. \nonumber\\
	&& + \frac{1}{2} \gamma^{\mu} \sum_{j = \pm}\lambda_{j}^{K} \left(A_{K^{*}}K^{*j}_{\mu} + B_{K^{*}}K^{*'j}_{\mu}\right) \nonumber\\
	&& + i \gamma^{5} \sum_{j = \pm} \lambda_{j}^{K} \left(A_{K}K^{j} + B_{K}K'^{j}\right) + \frac{1}{2} \gamma^{\mu} \lambda^{\omega} A_{\omega}\omega_{\mu} \nonumber\\
	&& \left. + \frac{1}{2} \gamma^{\mu} \lambda^{\phi} A_{\phi}\phi_{\mu} + \frac{1}{2} \gamma^{\mu} \lambda^{\rho} A_{\rho}\rho_{\mu} \right]q,
	\end{eqnarray}
	where $q$ and $\bar{q}$ are the $U(3)$ triplets of the u, d and s quark fields with the constituent masses $m_{u} \approx m_{d} = 280$~MeV, $m_{s} = 420$~MeV, the excited meson states are marked with prime. The masses $m_{u}$ and $m_{d}$ and the cutoff parameter (restricts the energy range in the integrals of the model) were obtained on the basis of the pion decay constant $F_{\pi}$ and $\rho$ meson decay constant $g_{\rho}$, which were calculated from the well-measured processes $\pi \to \mu \nu_{\mu}$ and $\rho(770) \to \pi \pi$. The mass $m_{s}$ was obtained on the basis of the kaon mass via the generalisation of the Gell-Mann-Oakes-Ranner formula to the case of strange mesons \cite{Volkov:1986zb}.
	
	The multipliers $A_{M}$ and $B_{M}$ were obtained as a result of diagonalisation of the Lagrangian with the initial unphysical ground and first radially excited states:
	\begin{eqnarray}
	\label{verteces}
	\label{coupling}
	A_{M} & = & \frac{1}{\sin(2\theta_{M}^{0})}\left[g_{M}\sin(\theta_{M} + \theta_{M}^{0}) +	g_{M}^{'}f_{M}(k_{\perp}^{2})\sin(\theta_{M} - \theta_{M}^{0})\right], \nonumber\\
	B_{M} & = & \frac{-1}{\sin(2\theta_{M}^{0})}\left[g_{M}\cos(\theta_{M} + \theta_{M}^{0}) +	g_{M}^{'}f_{M}(k_{\perp}^{2})\cos(\theta_{M} - \theta_{M}^{0})\right].
	\end{eqnarray}
	Here the index $M$ denotes an appropriate meson.
	
	The form factor $f\left(k_{\perp}^{2}\right) = \left(1 + d k_{\perp}^{2}\right)\Theta(\Lambda^{2} - k_{\perp}^2)$ describes the first radially excited mesons, where $\Lambda = 1.03$~GeV is the cutoff parameter, and the slope parameter $d$ depends only on the quark composition of the meson \cite{Volkov:2017arr}:
	\begin{eqnarray}
	&d_{uu} = -1.784 \times 10^{-6} \textrm{MeV}^{-2}, \quad d_{ss} = -1.737 \times 10^{-6}\textrm{MeV}^{-2},& \nonumber\\
	&d_{us} = -1.761 \times 10^{-6}\textrm{MeV}^{-2}.&
	\end{eqnarray}
	This parameter was obtained from the requirement of invariability of the quark condensate after including the radially excited	meson states.
	
	The transverse relative momentum of the inner quark-antiquark system can be represented as
	\begin{eqnarray}
	k_{\perp} = k - \frac{(kp) p}{p^2},
	\end{eqnarray}
	where $p$ is the meson momentum. In the rest system of a meson (see details in \cite{Volkov:2018fyv})
	\begin{eqnarray}
	k_{\perp} = (0, {\bf k}).
	\end{eqnarray}
	Therefore, this momentum may be used as a three-dimensional one.
	
	The parameter  $\theta_{M}$ is the mixing angle for the ground and the first radially excited mesons \cite{Volkov:1996fk,Volkov:2017arr}:
	\begin{eqnarray}
	\theta_{K_{1}} = 85.97^{\circ}, &\quad& \theta_{K^{*}} = 84.74^{\circ}, \nonumber\\
	\theta_{K} = 58.11^{\circ}, &\quad& \theta_{\omega} = 81.8^{\circ}, \nonumber\\
	\theta_{\phi} = 68.4^{\circ}. &&
	\end{eqnarray}
	Auxiliary values $\theta_{M}^{0}$ are included for convenience:
	\begin{eqnarray}
	\label{tetta0}
	&\sin\left(\theta_{M}^{0}\right) = \sqrt{\frac{1 + R_{M}}{2}},& \nonumber\\
	&R_{K_{1}} = R_{K^{*}} = \frac{I_{11}^{f_{us}}}{\sqrt{I_{11}I_{11}^{f^{2}_{us}}}}, \quad R_{K} = \frac{I_{11}^{f_{us}}}{\sqrt{Z_{K}I_{11}I_{11}^{f^{2}_{us}}}},& \nonumber\\
	&R_{\omega} = \frac{I_{20}^{f_{uu}}}{\sqrt{I_{20}I_{20}^{f^{2}_{uu}}}}, \quad R_{\phi} = \frac{I_{02}^{f_{ss}}}{\sqrt{I_{02}I_{02}^{f^{2}_{ss}}}},&
	\end{eqnarray}
	where
	\begin{eqnarray}
	\label{Zk}
	Z_{K} = \left[1 - \frac{3}{2}(m_{u} + m_{s})^{2}\left(\frac{\sin^{2}{\alpha}}{M^{2}_{K_{1(1270)}}} + \frac{\cos^{2}{\alpha}}{M^{2}_{K_{1(1400)}}}\right)\right]^{-1}
	\end{eqnarray}
	is an additional constant of renormalization appearing in $K - K_{1}$ transitions; $M_{K_{1}(1270)} = 1272 \pm 7$~MeV and $M_{K_{1}(1400)} = 1403 \pm 7$~MeV are the masses of the axial vector strange mesons \cite{Tanabashi:2018oca}. This form of the value $Z_{K}$ takes into account the mixing between the mesons $K_{1}(1270)$ and $K_{1}(1400)$ \cite{Volkov:2019yhy}.
	
	The integrals appearing in the quark loops as a result of renormalization of the Lagrangian are
	\begin{eqnarray}
	I_{l_{1}l_{2}}^{f^{n}} =
	-i\frac{N_{c}}{(2\pi)^{4}}\int\frac{f^{n}({\bf k}^{2})}{(m_{u}^{2} - k^2)^{l_{1}}(m_{s}^{2} - k^2)^{l_{2}}}		\mathrm{d}^{4}k.
	\end{eqnarray}
	
	Then
	\begin{eqnarray}
	\theta_{K_{1}}^{0} = \theta_{K^{*}}^{0} = 59.56^{\circ}, &\quad& \theta_{K}^{0} = 55.52^{\circ}, \nonumber\\
	\theta_{\omega}^{0} = 61.5^{\circ}, &\quad& \theta_{\phi}^{0} = 57.13^{\circ}. \nonumber\\
	\end{eqnarray}
	
	The coupling constants are:
	\begin{eqnarray}
	\label{coupling_constant}
	g_{K_{1}} = g_{K^{*}} = \left(\frac{2}{3}I_{11}\right)^{-1/2}, &\quad& g_{\omega} =  \left(\frac{2}{3}I_{20}\right)^{-1/2},  \nonumber\\
	g_{K} =  \left(\frac{4}{Z_{K}}I_{11}\right)^{-1/2}, &\quad& g_{\phi} =  \left(\frac{2}{3}I_{02}\right)^{-1/2}, \nonumber\\
	g_{K_{1}}^{'} = g_{K^{*}}^{'} = \left(\frac{2}{3}I_{11}^{f^{2}}\right)^{-1/2}, &\quad& g_{K}^{'} =  \left(4I_{11}^{f^{2}}\right)^{-1/2}, \nonumber\\
	g_{\omega}^{'} =  \left(\frac{2}{3}I_{20}^{f^{2}}\right)^{-1/2}, &\quad& g_{\phi}^{'} =  \left(\frac{2}{3}I_{02}^{f^{2}}\right)^{-1/2}.
	\end{eqnarray}
	They are the renormalisation constants of the quark-meson Lagrangian before diagonalization.
	
	In the Lagrangian (\ref{Lagrangian}), $K_{1}$ is the meson $K_{1A}$ defined in (\ref{K1AK1B}).

\section{The amplitudes of the processes in the extended NJL model}	
\begin{figure}[h]
		\center{\includegraphics[scale = 0.8]{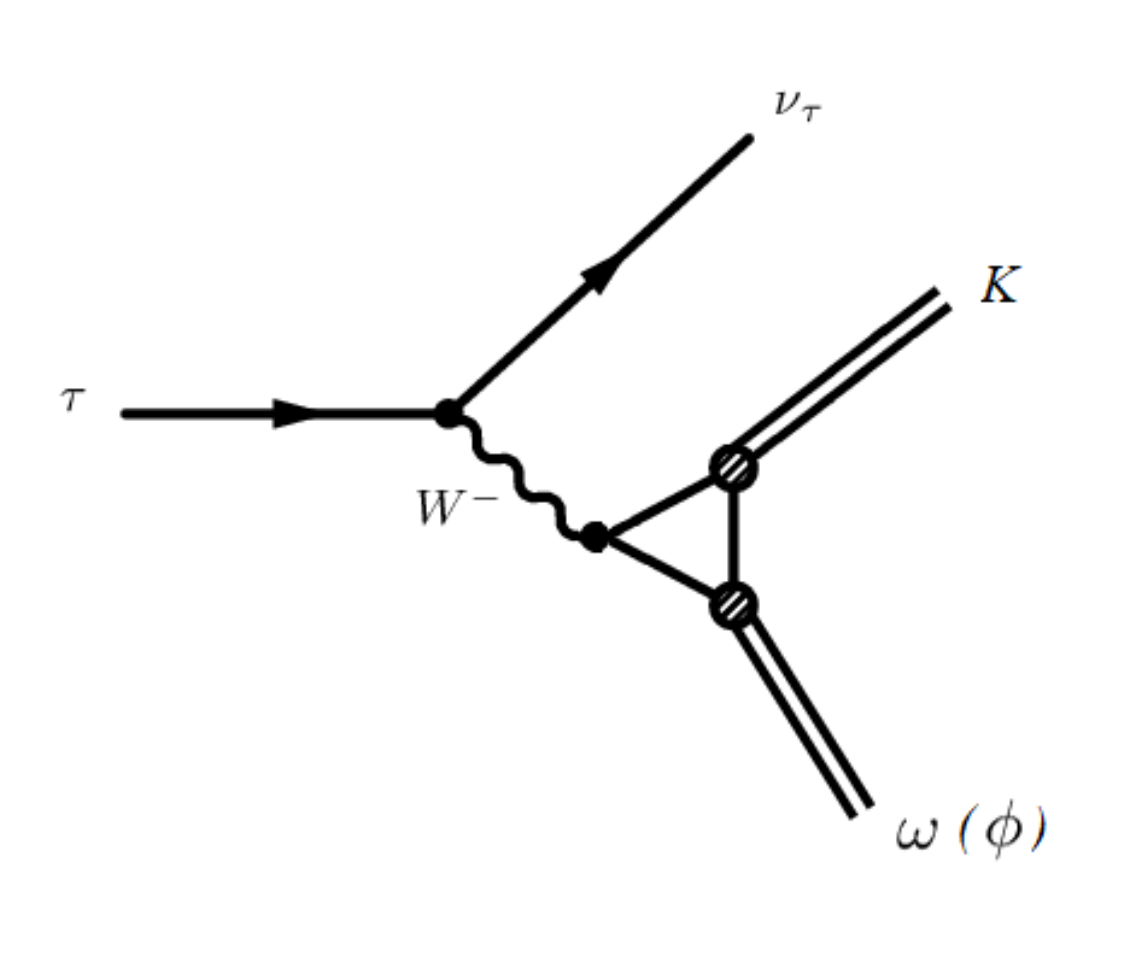}}
		\caption{The contact diagram.}
		\label{Contact}
	\end{figure}

\begin{figure}[h]
		\center{\includegraphics[scale = 1]{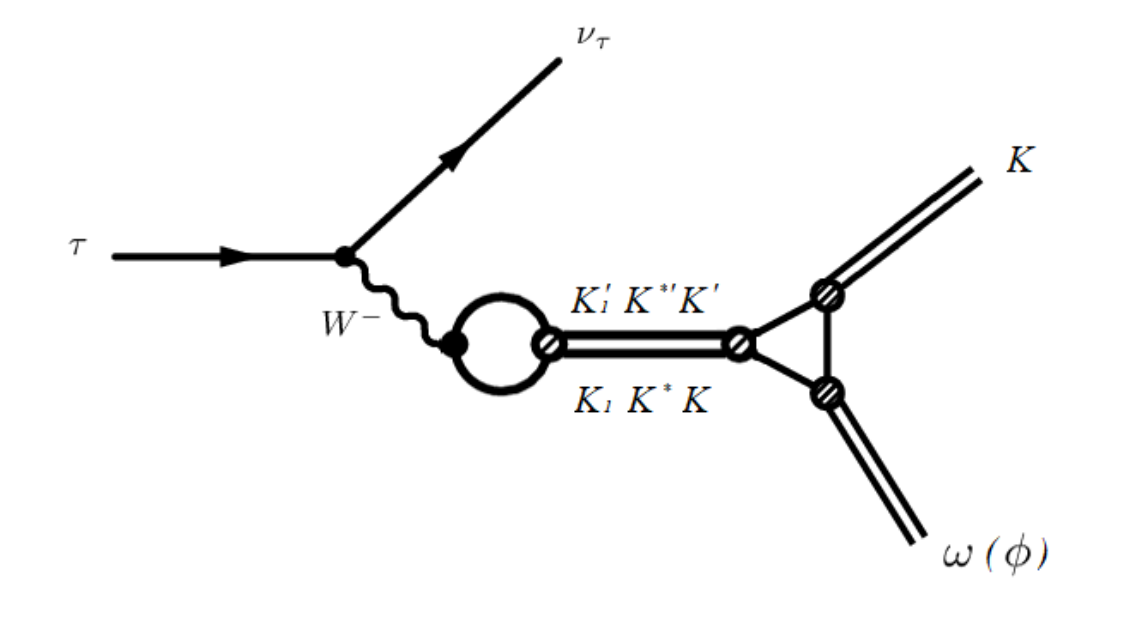}}
		\caption{The diagram with the intermediate mesons.}
		\label{Intermediate}
	\end{figure}

	The diagrams for the processes $\tau \to [\omega(782), \phi(1020)] K^{-} \nu_{\tau}$ are shown in Figs.~\ref{Contact},~\ref{Intermediate}.
	
	
	The amplitude of the process $\tau \to \omega (782) K \nu_{\tau}$ in the extended NJL model takes the form
	\begin{eqnarray}
	\mathcal{M} & = & -i G_{F} V_{us} L_{\mu} \left\{ \mathcal{M}_{c} + \mathcal{M}_{A} + \mathcal{M}_{V} + \mathcal{M}_{P} \right. \nonumber\\
	&& \left. + \mathcal{M}_{A^{'}} + \mathcal{M}_{V^{'}} + \mathcal{M}_{P^{'}} \right\}^{\mu\nu} e_{\nu}^{*}(p_{\omega}),
	\end{eqnarray}
	where $G_{F}$ is the Fermi constant, $V_{us}$ is the Cabbibo-Kobayashi-Maskawa matrix element, $L_{\mu}$ is the lepton current and $e_{\nu}^{*}(p_{\omega})$ is the polarization vector of the meson $\omega(782)$. The terms in the brackets describe the contributions from the contact diagram and from the diagrams with the intermediate axial vector, vector and pseudoscalar mesons in the ground and the first radially excited states:
	\begin{eqnarray}
	\label{amplitude}
	\mathcal{M}_{c}^{\mu\nu} & = & (m_{s} + m_{u}) I_{11}^{K\omega} g^{\mu\nu} + i 2 m_{u} \left[I_{21}^{K\omega} + (m_{s} - m_{u}) m_{u} I_{31}^{K\omega}\right] e^{\mu\nu\lambda\delta} p_{K\lambda} p_{\omega\delta}, \nonumber\\
	\mathcal{M}_{A}^{\mu\nu} & = & \frac{C_{K_{1}}}{g_{K_{1}}} (m_{s} + m_{u}) I_{11}^{K\omega K_{1}} \nonumber\\
	&& \times \left\{\left[g^{\mu\nu}\left[q^{2} - \frac{3}{2}(m_{s} + m_{u})^{2}\right] - q^{\mu}q^{\nu}\left[1 - \frac{3}{2}\frac{(m_{s} + m_{u})^{2}}{M_{K_{1(1270)}}^{2}}\right]\right]BW_{K_{1(1270)}}\sin^{2}{\alpha} \right. \nonumber\\
	&& \left. + \left[g^{\mu\nu}\left[q^{2} - \frac{3}{2}(m_{s} + m_{u})^{2}\right] - q^{\mu}q^{\nu}\left[1 - \frac{3}{2}\frac{(m_{s} + m_{u})^{2}}{M_{K_{1(1400)}}^{2}}\right]\right]BW_{K_{1(1400)}}\cos^{2}{\alpha}\right\}, \nonumber\\
	\mathcal{M}_{V}^{\mu\nu} & = & i 2 m_{u} \frac{C_{K^{*}}}{g_{K^{*}}} \left[I_{21}^{K\omega K^{*}} + (m_{s} - m_{u}) m_{u} I_{31}^{K\omega K^{*}}\right] \left[q^{2} - \frac{3}{2}(m_{s} - m_{u})^{2}\right] BW_{K^{*}} e^{\mu\nu\lambda\delta} p_{K\lambda} p_{\omega\delta}, \nonumber\\
	\mathcal{M}_{P}^{\mu\nu} & = & 2 (m_{s} + m_{u}) \frac{Z_{K}}{g_{K}} C_{K} I_{11}^{\omega K K} q^{\mu}q^{\nu} BW_{K}, \nonumber\\
	\mathcal{M}_{A^{'}}^{\mu\nu} & = & \frac{C_{K_{1}}^{'}}{g_{K_{1}}} (m_{s} + m_{u}) I_{11}^{K\omega K_{1}^{'}} \left\{g^{\mu\nu}\left[q^{2} - \frac{3}{2}(m_{s} + m_{u})^{2}\right] - q^{\mu}q^{\nu}\left[1 - \frac{3}{2}\frac{(m_{s} + m_{u})^{2}}{M_{K_{1}'}^{2}}\right]\right\} BW_{K_{1}^{'}}, \nonumber\\
	\mathcal{M}_{V^{'}}^{\mu\nu} & = & i 2 m_{u} \frac{C_{K^{*}}^{'}}{g_{K^{*}}} \left[I_{21}^{K\omega K^{*'}} + (m_{s} - m_{u}) m_{u} I_{31}^{K\omega K^{*'}}\right] \left[q^{2} - \frac{3}{2}(m_{s} - m_{u})^{2}\right] BW_{K^{*'}}  e^{\mu\nu\lambda\delta} p_{K\lambda}p_{\omega\delta}, \nonumber\\
	\mathcal{M}_{P^{'}}^{\mu\nu} & = & 2 (m_{s} + m_{u}) \frac{Z_{K}}{g_{K}} C_{K}^{'} I_{11}^{K\omega K^{'}} q^{\mu}q^{\nu} BW_{K}.
	\end{eqnarray}
	Two terms in the contact contribution describe the axial vector and the vector parts of the contact diagram.
	
	The Breit-Wigner propagator takes the following form:
	\begin{eqnarray}
	BW_{M} = \frac{1}{M_{M}^{2} - q^{2} - i\sqrt{q^{2}}\Gamma_{M}}.
	\end{eqnarray}
	
	The constants $C_{M}$ and $C_{M}^{'}$ appear in the quark loops of the $W$-boson transition into the intermediate meson in the extended NJL model:
	\begin{eqnarray}
	C_{M} & = & \frac{1}{\sin\left(2\theta_{M}^{0}\right)}\left[\sin\left(\theta_{M} + \theta_{M}^{0}\right) +	R_{M}\sin\left(\theta_{M} - \theta_{M}^{0}\right)\right], \nonumber \\
	C_{M}^{'} & = & \frac{-1}{\sin\left(2\theta_{M}^{0}\right)}\left[\cos\left(\theta_{M} + \theta_{M}^{0}\right) + R_{M}\cos\left(\theta_{M} - \theta_{M}^{0}\right)\right].
	\end{eqnarray}
	The values $R$ are defined in (\ref{tetta0}).
	
	The integrals with the vertices from the Lagrangian in the numerator, which were also used in the amplitude, take the form:
	\begin{eqnarray}
	\label{DiffIntegral}
	I_{n_{1} n_{2}}^{M, \dots, M^{'}, \dots} & = &
	-i\frac{N_{c}}{(2\pi)^{4}}\int\frac{A_{M} \dots B_{M} \dots}{(m_{u}^{2} - k^2)^{n_{1}}(m_{s}^{2} - k^2)^{n_{2}}} \Theta(\Lambda^{2} - {\bf k}^2)	\mathrm{d}^{4}k,
	\end{eqnarray}
	where $A_{M}, B_{M}$ are defined in (\ref{verteces}).
	
	The amplitude for the process $\tau \to \phi(1020) K^{-} \nu_{\tau}$ differs from the amplitude written above by the factor $2$, by the replacement of the appropriate vertices in (\ref{DiffIntegral}) and by the replacement of the masses of $u$ and $s$ quarks in the vector channel and in the vector part of the contact channel.
	
\section{Numerical estimations}
	There are different ways to choose the mixing angle $\alpha$ in (\ref{K1AK1B}). In the Particle Data Group (PDG), the value $\alpha = 45^{\circ}$ is indicated \cite{Tanabashi:2018oca}. However, in \cite{Suzuki:1993yc}, it is shown that this angle can also take the values $\alpha = 57^{\circ}$ and $\alpha = 33^{\circ}$. In the work \cite{Volkov:1984fr}, the value $\alpha = 57^{\circ}$ was obtained. In our recent work \cite{Volkov:2019yhy}, it has been shown that the value $\alpha = 57^{\circ}$ leads to the best agreement of the kaon mass, which is calculated using the generalisation of the Gell-Mann-Oaks-Renner formula to the kaons, with the experimental value.
	
	In this paper, we demonstrate the results for two possible values: $\alpha = 57^{\circ}$ and $\alpha = 45^{\circ}$.
	
	The branching fractions of the considered processes are shown in Table~\ref{tab:1}. 
	\begin{table}
		\caption{The branching fractions of the processes $\tau \to \omega(782) K \nu_{\tau}$ and $\tau \to \phi(1020) K \nu_{\tau}$ for different values of the angle $\alpha$. The contributions from various channels are shown in different lines. Lines $W_{A}$ and $W_{V}$ correspond to the axial vector and vector parts of the contact term. The line Ground contains the summary results for the contributions of all ground intermediate mesons and contact diagram. The line Excited contains the summary results for the contributions of all excited intermediate mesons.}
		\label{tab:1}       
		\begin{center}
		\begin{tabular}{ccccc}
			\hline\noalign{\smallskip}
			&\multicolumn{4}{c}{Br $(\times 10^{-4})$}\\
			\noalign{\smallskip}\hline\noalign{\smallskip}
			&\multicolumn{2}{c}{$\tau \to \omega(782) K \nu_{\tau}$}&\multicolumn{2}{c}{$\tau \to \phi(1020) K \nu_{\tau}$}\\
			\noalign{\smallskip}\hline\noalign{\smallskip}
			& $\alpha = 57^{\circ}$ & $\alpha = 45^{\circ}$ & $\alpha = 57^{\circ}$ & $\alpha = 45^{\circ}$ \\
			\noalign{\smallskip}\hline\noalign{\smallskip}
			W$_{A}$	 	 & 0.54 				& 0.52				   & 2.02				  & 1.97  \\
			A        	 & 4.73 				& 4.47 				   & 7.49				  & 8.59  \\
			W$_{A}$ + A	 & 3.08 				& 3.27				   & 2.12				  & 2.94  \\
			W$_{V}$	 	 & 0.66 				& 0.64				   & 0.84				  & 0.82  \\
			V        	 & 1.99 				& 1.94         		   & 0.89				  & 0.86	\\
			W$_{V}$ + V	 & 0.37 				& 0.36         		   & 2.6 $\times 10^{-3}$ & 2.5 $\times 10^{-3}$	\\
			P        	 & 0.57 				& 0.53 				   & 0.66				  & 0.61    \\
			Ground		 & 3.83 				& 3.96 				   & 2.57				  & 3.34    \\
			A$^{'}$   	 & 5.4 $\times 10^{-3}$ & 5.2 $\times 10^{-3}$ & 0.74				  & 0.72  \\
			V$^{'}$   	 & 0.31 				& 0.3 				   & 18.1 $\times 10^{-3}$& 17.7 $\times 10^{-3}$  \\
			P$^{'}$   	 & 8.6 $\times 10^{-4}$ & 6.1 $\times 10^{-4}$ & 1.6 $\times 10^{-4}$ & 1.2 $\times 10^{-4}$ \\
			Excited 	 & 0.32 				& 0.31 				   & 0.76				  & 0.74  \\
			\noalign{\smallskip}\hline\noalign{\smallskip}
			Total      	 & 3.79 				& 3.95 				   & 3.15				  & 4.04    \\
			\noalign{\smallskip}\hline\noalign{\smallskip}
			Exp      	 & \multicolumn{2}{c}{$4.1 \pm 0.9$ \cite{Tanabashi:2018oca}}&\multicolumn{2}{c}{$4.4 \pm 1.6$ \cite{Tanabashi:2018oca}}         \\
			& \multicolumn{2}{c}{}&\multicolumn{2}{c}{$4.05 \pm 0.51$ \cite{Inami:2006vd}}         \\
			& \multicolumn{2}{c}{}&\multicolumn{2}{c}{$3.39 \pm 0.48$ \cite{Aubert:2007mh}}         \\
			\noalign{\smallskip}\hline
		\end{tabular}
		\end{center}
	\end{table}
	
\begin{figure}[h]
		\center{\includegraphics[scale = 0.9]{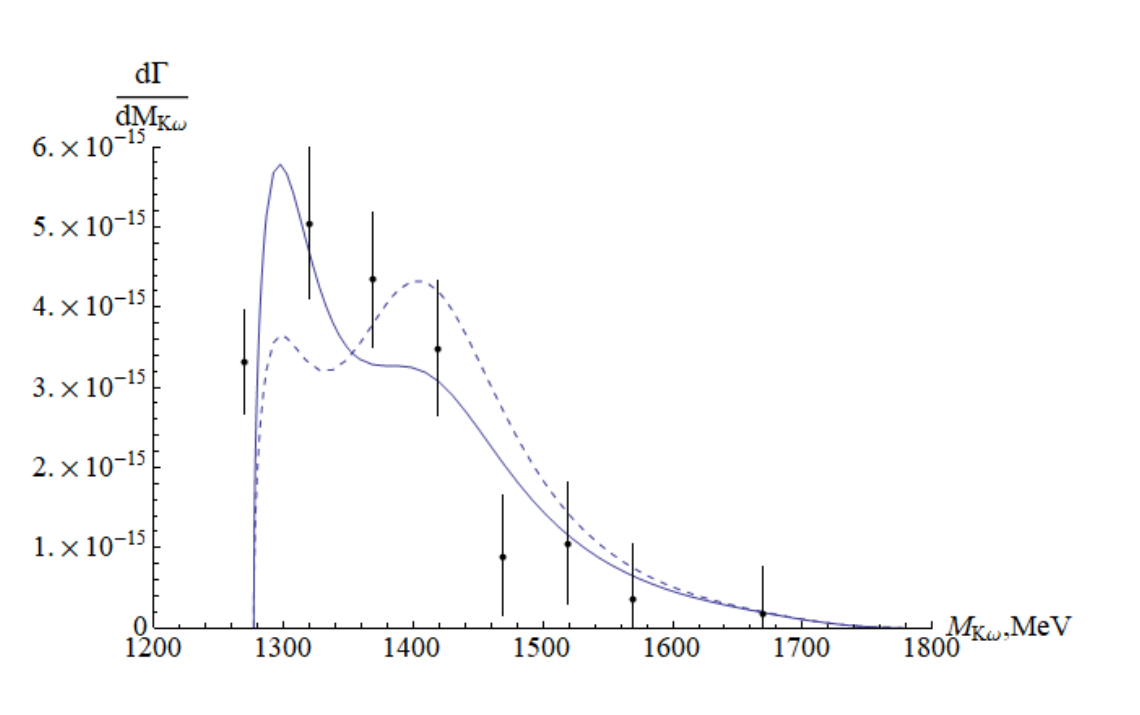}}
		\caption{The differential decay width for the process $\tau \to \omega(782) K \nu_{\tau}$. The solid line corresponds to the case $\alpha = 57^{\circ}$, the dashed line corresponds to the case $\alpha = 45^{\circ}$, the experimental points are taken from \cite{Arms:2005qg}.}
		\label{omega}
	\end{figure}

\begin{figure}[h]
		\center{\includegraphics[scale = 0.9]{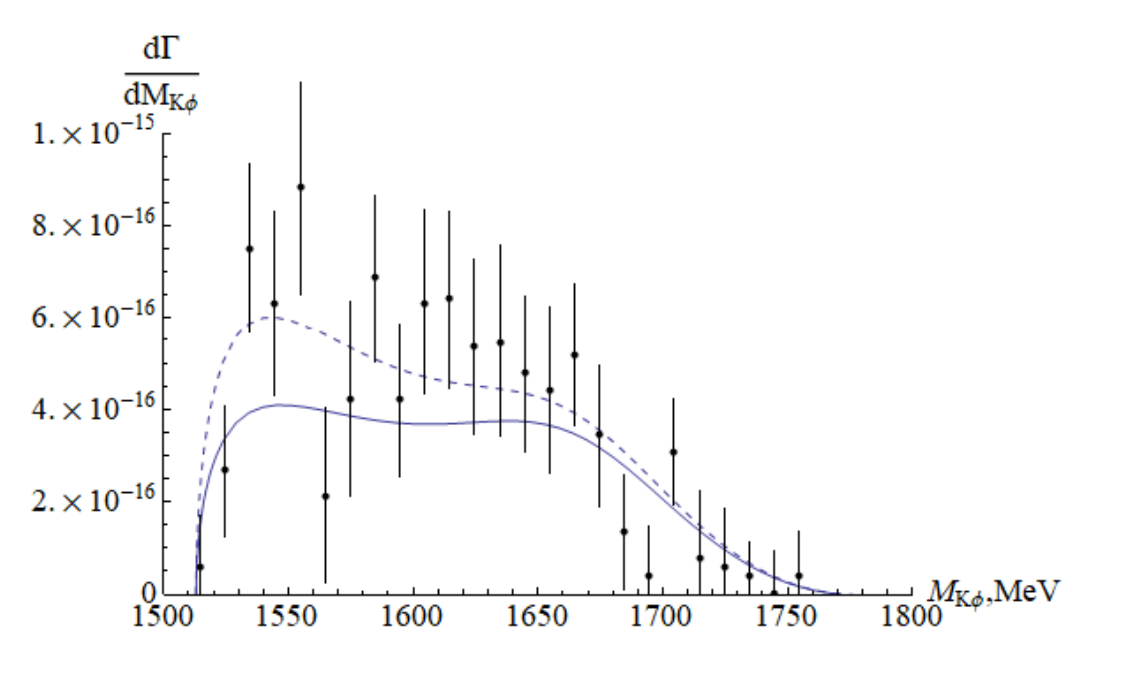}}
		\caption{The differential decay width for the process $\tau \to \phi(1020) K \nu_{\tau}$. The solid line corresponds to the case $\alpha = 57^{\circ}$, the dashed line corresponds to the case $\alpha = 45^{\circ}$, the experimental points are taken from \cite{Inami:2006vd}.}
		\label{phy}
	\end{figure}

	The comparison of the differential decay width for our processes with the experimental data are shown in Figs.~\ref{omega},~\ref{phy}.
	
	As one can see from Table~\ref{tab:1}, in all cases the ground axial vector channel (diagrams with the intermediate mesons $K_{1}(1270)$ and $K_{1}(1400)$) gives the main contribution. The interference between the ground axial vector or vector channels and the corresponding parts of the contact terms is always negative due to the negative value of the Breit-Wigner propagator for most of the energy range. 
	
	The mixing angle  $\alpha$ affects not only the ground axial vector channel but also all others, because the coupling constant of the final kaon with quarks contains the constant $Z_{K}$ that depends on this angle (see (\ref{coupling}), (\ref{Zk}), (\ref{coupling_constant})). Transition from the angle $\alpha = 57^{\circ}$ to $\alpha = 45^{\circ}$ reduces all contributions except the contribution of the ground axial vector mesons in the process $\tau \to \phi(1020) K \nu_{\tau}$. However, in the process $\tau \to \omega(782) K \nu_{\tau}$, despite the decrease of the contribution of the ground axial vector channel and axial vector part of the contact channel, their summary contribution increased. This may be explained by the decrease of the negative interference between them. In this way, for both processes the final result for $\alpha = 45^{\circ}$ is larger than for $\alpha = 57^{\circ}$.
	
	The interference between the vector channels or vector parts of the contact diagrams and other channels is equal to zero. This is because the indices $\mu\nu$ in (\ref{amplitude}) are antisymmetric in the first ones and symmetric in the other channels. Thus, one can see that the interference between the axial vector and the pseudoscalar channel is always negative.
	
	As one can see from formula (\ref{amplitude}), the term corresponding to the diagram with the intermediate meson $K_{1}(1400)$ is proportional to $\cos^{2}{\alpha}$ and should play a more important role in the case $\alpha = 45^{\circ}$ than in the case $\alpha = 57^{\circ}$. This is the cause of the increase of the second peak in the dashed line of the differential decay width for the process $\tau \to \omega(782) K \nu_{\tau}$ (see Fig.~\ref{omega}) and the first peak for the process $\tau \to \phi(1020) K \nu_{\tau}$(see Fig.~\ref{phy}). The second peak in Fig.~\ref{phy} corresponds to the contribution from the meson $K_{1}(1650)$ which plays a significant role in this process compared to the process $\tau \to \omega(782) K \nu_{\tau}$.
	
\section{Discussion and Conclusion}
    Our calculations show that taking into account the mixing of two axial vector mesons plays an important role. It leads to the physical states $K_{1}(1270)$ and $K_{1}(1400)$. It is interesting to compare the results obtained with different values of the mixing angle $\alpha$. The obtained branching fractions of the process $\tau \to \omega(782) K \nu_{\tau}$ for $\alpha = 57^{\circ}$ and $\alpha = 45^{\circ}$ are within the errors of the experimental values. The obtained branching fractions of the process $\tau \to \phi(1020) K \nu_{\tau}$ with different mixing angles agree with different experimental results. It is interesting to note that the value $\alpha = 57^{\circ}$ leads to better agreement of the form of the invariant mass distribution of the process $\tau \to \omega(782) K \nu_{\tau}$ with the experimental points. However, the invariant mass distribution of the decay $\tau \to \phi(1020) K \nu_{\tau}$ with $\alpha = 45^{\circ}$ is in better agreement with the experimental data. In any case, our results are consistent with the experimental data with allowance for the precision of the model, which was estimated in Introduction and is expected to be near $17\%$. Further experimental studies can help better understand the effects of mixing axial vector mesons and give us additional restrictions for the angle $\alpha$.
    
    The question concerning the nature of the intermediate axial vector meson is not completely clear. As was mentioned in Introduction, our approach considers the meson $K_{1}(1270)$ as a $q\bar{q}$ resounance. However, this meson can have a more complicated structure. It may have a two-pole nature \cite{Roca:2005nm,Wang:2019mph} or even be a tetraquark meson state. It would be interesting to continue the research of this question.
    
\section*{Acknowlegments}
	The authors are grateful to A. B. Arbuzov for useful discussions; this work is supported by the JINR grant for young scientists and specialists No. 19-302-06.

\end{document}